\documentclass[conference]{IEEEtran}
\IEEEoverridecommandlockouts

\usepackage{cite}

\ifCLASSINFOpdf
  \usepackage[pdftex]{graphicx}
  \DeclareGraphicsExtensions{.pdf,.jpeg,.png}
\else
  \usepackage[dvips]{graphicx}
  \DeclareGraphicsExtensions{.eps}
\fi
\usepackage{amsmath}
\usepackage{array}

\ifCLASSOPTIONcompsoc
  \usepackage[caption=false,font=normalsize,labelfont=sf,textfont=sf]{subfig}
\else
  \usepackage[caption=false,font=footnotesize]{subfig}
\fi
\usepackage{url}

\usepackage{comment}

\newcommand{\mxsrc}{\ensuremath{\mathit{src}}}
\newcommand{\mxsnk}{\ensuremath{\mathit{snk}}}
\newcommand{\mxin}{\ensuremath{\mathit{in}}}
\newcommand{\mxout}{\ensuremath{\mathit{out}}}
\newcommand{\mxblocks}{\ensuremath{\mathit{blks}}}
\newcommand{\mxpassive}{\ensuremath{\mathit{pssv}}}
\newcommand{\mxactive}{\ensuremath{\mathit{actv}}}
\newcommand{\mxpred}{\ensuremath{\mathit{pred}}}
\newcommand{\mxsucc}{\ensuremath{\mathit{succ}}}

\newcommand{\mxinpair}{\ensuremath{\kappa_{i}}}
\newcommand{\mxallin}{\ensuremath{\mathcal{K}_{i}}}
\newcommand{\mxallout}{\ensuremath{\mathcal{K}_{o}}}
\newcommand{\mxoutpair}{\ensuremath{\kappa_{o}}}
\newcommand{\mxallactors}{\ensuremath{\mathcal{A}}}

\begin{document}
%
\title{Generalized Graph Connections for Dataflow Modeling of DSP Applications
\thanks{This is a pre-publication version of a paper that has been
accepted for publication in the 2018 IEEE International Workshop on
Signal Processing Systems. The official/final version of the paper
will be posted on IEEE Xplore.}}

\author{\IEEEauthorblockN{Yanzhou Liu\IEEEauthorrefmark{1},
                          Lee Barford\IEEEauthorrefmark{2},
                          Shuvra S. Bhattacharyya\IEEEauthorrefmark{1}\IEEEauthorrefmark{3}}
\IEEEauthorblockA{\IEEEauthorrefmark{1}Department of Electrical and Computer
Engineering, University of Maryland, College Park, MD 20742, USA}
\IEEEauthorblockA{\IEEEauthorrefmark{2}Keysight Laboratories, Keysight Technologies, 
Reno, NV, USA}
\IEEEauthorblockA{\IEEEauthorrefmark{3}Department of Pervasive Computing Tampere University 
of Technology, Tampere, Finland}
}


\maketitle


\begin{abstract}
In dataflow representations for signal processing systems,
applications are represented as directed graphs in
which vertices represent computations and edges
correspond to buffers that store data as it passes
between computations. The buffers are single-input, single-output
components that manage data in a first-in, first-out (FIFO) fashion.
In this paper, we generalize the concept of dataflow buffers
with a concept called ``passive blocks''. Like dataflow buffers, passive
blocks are used to store data during the intervals between
its generation by producing actors, and its use by consuming
actors. However, passive blocks can have multiple inputs and
multiple outputs, and can incorporate operations on and
rearrangements of the
stored data subject to certain constraints. We define
a form of flowgraph representation that is based
on replacing dataflow edges with the proposed concept of passive blocks.
We present a structured design methodology for 
utilizing this new form of signal processing flowgraph,
and demonstrate its utility in improving memory management efficiency,
and execution time performance.

\end{abstract}


%
\IEEEpeerreviewmaketitle


\section{Introduction}
\label{sec:intro}

Dataflow modeling is widely used in design processes and tools
for signal processing systems. In this form of modeling,
applications are represented as directed graphs, called {\em
dataflow graphs},
in which vertices ({\em actors}) represent
discrete computations that are executed iteratively
({\em fire}) to process semi-infinite streams of input
data. Each edge $e = (x, y)$ in a dataflow graph represents
a logical communication channel between actors $x$ and $y$.
More specifically, each $e = (x, y)$ represents
a first-in, first-out (FIFO) buffer that stores data
during the period between its production by actor $x$ and
its consumption by actor $y$. Actors can be fired
when certain conditions, referred to as {\em firing rules},
are satisfied~\cite{lee1995x1}. 

Dataflow modeling has proven to be of great utility in the design and
implementation of signal processing systems for various reasons, including its
provisions for ensuring determinacy, support for exploiting parallelism, and
capability for exposing high-level application structure that is useful for
many kinds of design optimization beyond those associated with exploiting
parallelism~\cite{bhat2013x4}.

\begin{figure}[!t]
\centering
\subfloat[]{\includegraphics[width=1.2in]{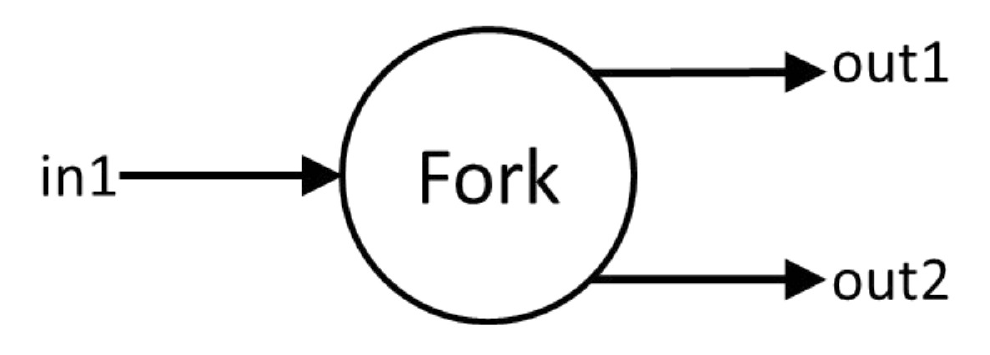}%
\label{fig:fork}}
\hfil
\subfloat[]{\includegraphics[width=2.1in]{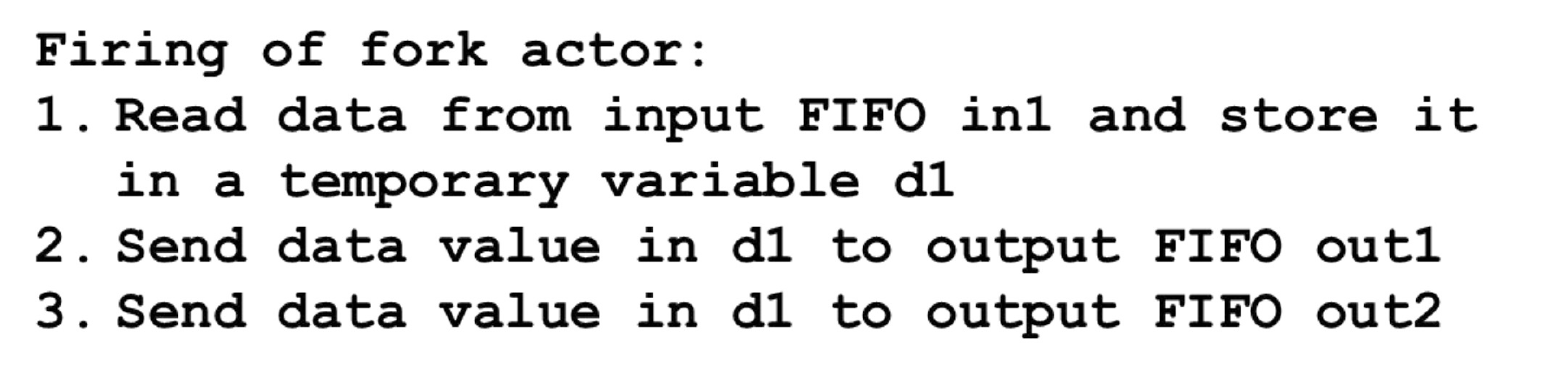}%
\label{fig:pseudo}}
\hfil
\subfloat[]{\includegraphics[scale=0.25]{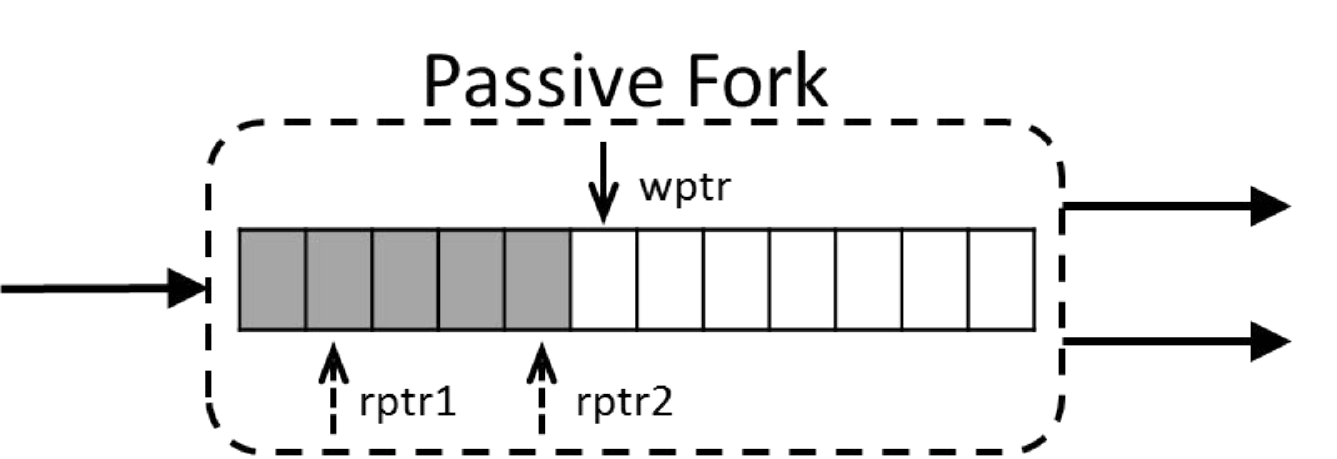}%
\label{fig:passive-fork}}
\hfil
\subfloat[]{\includegraphics[scale=0.25]{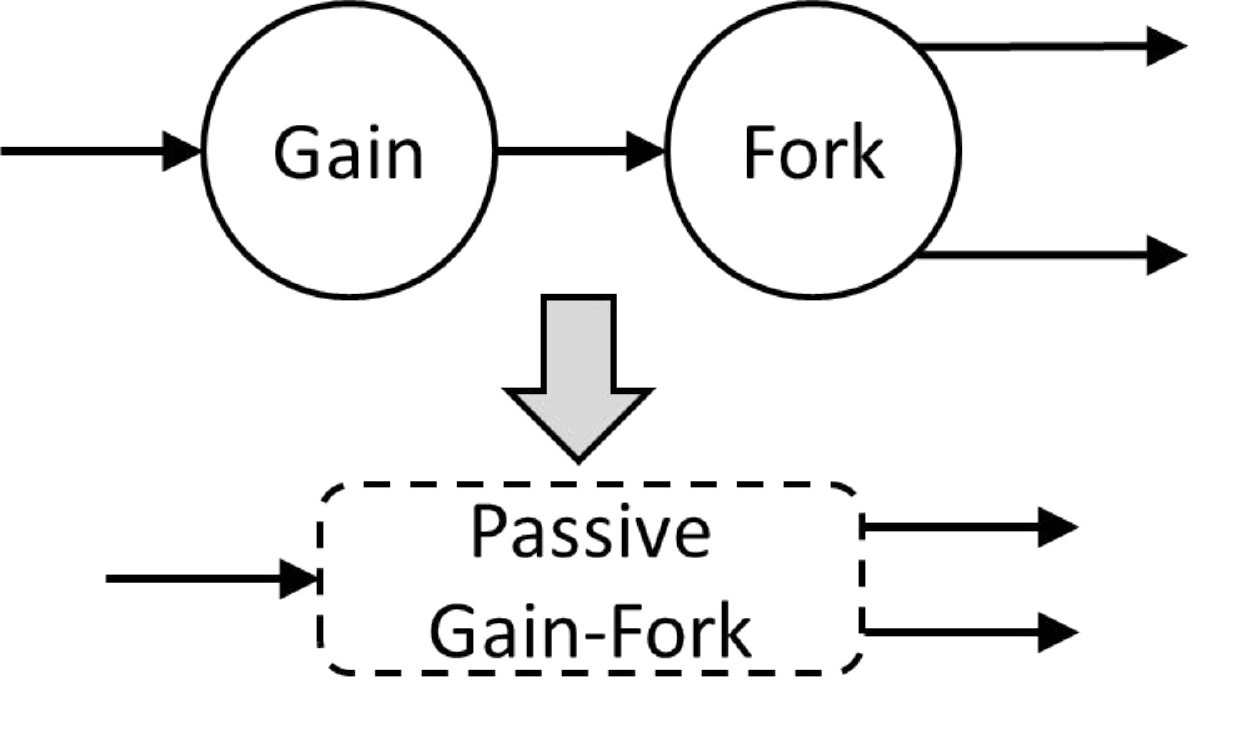}%
\label{fig:passive-gainfork}}
\caption{(a) Fork actor. (b) Pseudocode fragment for the fork 
actor. (c) Passive version of fork actor; wptr, rptr1, and rptr2 
show possible positions of the write pointer and the two read pointers. (d) Cascade of gain and fork actors.}
\label{fig:demo}
\end{figure}

A limitation of signal processing dataflow representations, however, is
that they are inefficient in describing inter-actor communication patterns
that depart from the simple 
single-input, single-output (SISO) interface and FIFO behavior
that are defined for dataflow edges. As a canonical example
of this kind of inefficiency, consider the
{\em fork} actor illustrated in Figure~\ref{fig:fork}.
This is a synchronous dataflow (SDF)~\cite{lee1987x1} actor
that consumes a single token $t$ and produces two tokens --- one
on each output edge --- on each firing. The values of the
two tokens that are produced are identical to the value of the
input token $t$. Thus, this actor can be viewed as providing
a kind of broadcast functionality.

Figure~\ref{fig:pseudo} shows a pseudocode fragment for 
the fork actor. From this pseudocode, we can see that 
there is overhead of copying the value of the input token
to each of the outputs. This overhead in general
includes a run-time cost as well as a cost in terms
of increased memory requirements.  The overhead is required under
a pure dataflow interpretation since the input token must be replicated
on each of the two output edges (FIFOs).

The functionality of the fork actor can be realized more efficiently if we
abandon this pure dataflow interpretation, and implement the actor instead
using the one-input, two-output component illustrated in
Figure~\ref{fig:passive-fork}. This component, which we refer to here as a {\em
passive fork}, is not {\em fired} as a dataflow actor is.  Instead, the
component operates in a manner similar to a typical FIFO implementation, where
a buffer is associated with the component, and tokens are written to and read
from the buffer using write and read pointers, respectively. However, the
passive fork has {\em two} read pointers --- one corresponding to each output
edge of the fork actor --- instead of the single read pointer that would be
used in a FIFO. In effect, we have transformed the fork actor, which operates
in an ``active'' manner (by firing) into an passive component, which is used by
writing to and reading from the component's ports. 

A more powerful form of this ``active-to-passive'' conversion is
illustrated on Figure~\ref{fig:passive-gainfork}, which shows a gain actor 
that is connected
at the input of the fork actor. This gain actor corresponds to a constant
multiplication, where the constant factor $k$ is a parameter of the actor. The
gain together with the fork can be replaced by a single passive component. This
component is similar to the passive fork actor, except that when a value is
written into the buffer, it is multiplied by $k$ before being stored. 
In this paper, we generalize this process of converting certain kinds of actors
into passive components, which achieve equivalent functionality through
read/write interfaces rather than through the mechanism of being fired.  This
generalization leads to a powerful new design methodology in which passive
components of arbitrary complexity can be designed to provide streamlined
functionality for actors or subgraphs that are more efficiently realized with
internal buffers and read/write interfaces. When dataflow graphs are
transformed to incorporate such passive components, we refer to the resulting
graphs as {\em Passive-Active Flow Graphs} ({\em PAFGs}).
A central objective of this paper is to introduce PAFGs as a useful
new representation for model-based design of signal processing systems.

\section{Related Work}
\label{sec:related}

Many researchers have investigated efficient buffer memory management in
dataflow graphs (e.g., see~\cite{fisc2007x1, oh2002x1, stui2006x1,
bhat2013x4}). Bhattacharyya and Lee discussed the concept that certain actors,
such as the fork actor described above, can be implemented more efficiently by
deviating from pure dataflow semantics~\cite{bhat1992x1}.  However, this
earlier work did not propose any approach for integrating such deviations
systematically into the modeling framework. In this new work, we develop such a
systematic approach based on the novel abstraction of PAFGs.

Perhaps the most closely related form of dataflow memory management
optimization to what we develop in this paper is {\em buffer merging}, which
involves mapping subsets of input and output buffers of a given actor to a
common memory space (e.g., see~\cite{murt2004x1, desn2015x1}).  Like the method
of~\cite{desn2015x1}, the PAFG approach allows for memory sharing across
arbitrary numbers of input and output buffers for a given actor. Similarly,
like the method of~\cite{murt2004x1}, the PAFG approach does not involve
expansion to a single rate rate graph, which can be costly in terms of compiler
memory requirements and time complexity for highly multirate applications
(e.g., see~\cite{hsu2007x1}). In this sense, the PAFG approach provides a novel
combination of useful features in the two previously developed buffer merging
approaches described above. Additionally, while the methods
of~\cite{murt2004x1, desn2015x1} are limited to SDF graphs, the PAFG approach
is not restricted to any specific form of dataflow. For example, Boolean
dataflow switch and select actors~\cite{buck1993x1} can be formulated as
optimized PAFG components using the same methodology that is presented in this
paper. Applicability
beyond SDF is also a distinguishing point compared to the 
abstraction of deterministic SDF with shared FIFOs (DSSF)~\cite{trip2013x1}.

While there are significant differences between buffer merging, DSSF, and
PAFG-based memory management, investigating and exploiting complementary
relationships among the different approaches is an interesting direction for
future work.



\section{PAFG Representations}
\label{sec:flowgraphs}


In this section, we develop in detail the PAFG model of computation.  In this
work, PAFGs are derived from dataflow graphs, and are intended as intermediate
representations or implementation architectures for dataflow {\em application
graphs} (dataflow models of signal processing applications).  For concreteness,
we develop the concepts of PAFGs here in the context of core functional
dataflow (CFDF) as the application graph model; however, the concepts are not
specific to CFDF and can be adapted to other forms of dataflow.  CFDF is a
highly expressive model that can be used to represent other well-known dataflow
models, including synchronous, cyclo-static, and Boolean
dataflow~\cite{plis2008x4}. CFDF is the model that underlies the lightweight
dataflow environment (LIDE) tool~\cite{lin2017x1}, which we use for our
experiments in Section~\ref{sec:experiments}.


We first define some notation that will be useful throughout the remainder of
this paper. Given an edge $e$ in a directed graph, we denote the source and
sink vertices of $e$ by $\mxsrc(e)$ and $\mxsnk(e)$, respectively. A {\em
self-loop} is an edge whose source and sink vertices are identical.  In the
remainder of this paper, we consider only directed graphs that do not contain
self-loops. Self-loops can be incorporated easily into the methods
developed in this paper; we omit the details due to space limitations.

Given an edge $e$, we say that $\mxsrc(e)$ is a {\em predecessor} of
$\mxsnk(e)$, $\mxsnk(e)$ is a {\em successor} of $\mxsrc(e)$, and $\mxsrc(e)$
and $\mxsnk(e)$ are {\em adjacent} vertices.  The sets of all predecessors and
successors of a vertex $v$ in a given graph are denoted by $\mxpred(v)$ and
$\mxsucc(v)$, respectively. The sets of all input edges and output edges
of $v$ are denoted by $\mxin(v)$ and $\mxout(v)$, respectively.

We refer to PAFG vertices as {\em blocks}.
In a dataflow graph, vertices correspond to computational
modules, and edges correspond to SISO
buffers between the modules. In contrast, in a PAFG,
both computational modules and buffers are represented as
vertices, and edges represent connections between 
computational modules and buffers. Additionally, PAFG
buffers are not restricted to SISO interfaces --- they can
have multiple inputs, multiple outputs, or both.
A third distinguishing characteristic of the PAFGs
that we are interested
in this paper is that they are bipartite graphs. We define this bipartite
characteristic precisely in Section~\ref{sec:alternating}.

For conciseness and clarity, we assume that dataflow graphs and PAFGs are
directed graphs rather than multigraphs (which can contain multiple edges
directed in the same direction and between the same pair of vertices).
The adaptation of the PAFG model to multigraphs can be readily
achieved when implementing the model.

We refer to an ordered pair of actors $(x_d, y_d)$ as a {\em dataflow
pair} and an ordered pair of PAFG blocks $(x_b, y_b)$
as a {\em PAFG pair}.

The PAFGs that we are concerned with in this paper are derived from
corresponding dataflow graphs (application graphs).  We elaborate on the process of deriving a
PAFG from a dataflow graph in Section~\ref{sec:direct}. This derivation process
places blocks in a PAFG $F$ in correspondence with actors or edges in the
dataflow graph from which $F$ was derived.  A {\em simple passive buffer}  is a
PAFG block that corresponds in this way to an edge in some dataflow graph.
A PAFG block that is not a simple passive buffer is referred to as a {\em
non-simple block}. We often refer to simple passive buffers as {\em simple
blocks}.

\subsection{PAFG Blocks}
\label{sec:components}

A PAFG block is either a {\em passive} block or an {\em active block}.
The distinction between these two types was motivated intuitively in
Section~\ref{sec:intro}. More precisely, an active block corresponds
to an application graph actor that is used in the usual way ---
that is, through interfaces that are associated with firing the actor
and (if available) for testing fireability. In CFDF, these are
referred to as {\em invoke} and {\em enable} interfaces, respectively~\cite{plis2008x4}. In contrast, a passive block is used through read/write interfaces,
as illustrated by the passive fork example in Section~\ref{sec:intro}.

Given an application graph $G$, we assume that implementations of
the actors in $G$ are available in an {\em actor library}.
We assume that each actor in $G$ has one
active implementation (with enable/invoke interfaces) in the library,
and that it may or may not have a passive implementation
(with read/write interfaces). We refer to an
actor $A$ as a {\em buffer actor} if it has a passive implementation;
otherwise, we refer to $A$ as a computational actor.
Thus, only buffer actors can be placed in correspondence
with passive blocks.


Like active blocks, non-simple passive blocks correspond to actors.  However,
they are used (executed) in a different way --- again, as illustrated by the
difference between the active (``standard'') and passive versions of the fork
actor in Section~\ref{sec:intro}.  A non-simple passive block should implement
the same input/output behavior as its corresponding actor --- that is, it
should perform the same mapping from input streams into output streams. For
background on the interpretation of actors as mappings from input streams to
output streams, we refer the reader to~\cite{lee1995x1}.  In this work, we
assume that unit testing processes are used to validate such ``mapping
equivalence'' between passive blocks and their corresponding actors. For
background on synergies between unit testing and dataflow-based design
processes, we refer the reader to~\cite{lin2017x1}.  We envision as an
interesting area for future work the automation of the equivalence checking
process between active and passive implementations of the same buffer actor.

A block in a PAFG is either a {\em computational block} or
a {\em buffer block}. The computational/buffer 
dichotomy is another relevant way to distinguish between
blocks in addition to the active/passive and simple/non-simple dichotomies.
All computational blocks are active blocks.
However, buffer blocks can in general be either passive or active. 


\subsection{Coordination Functions and Alternating PAFGs}

When deriving a PAFG, each buffer block needs to be designated as being an
active or passive buffer. An active buffer is executed like any other actor
(using enable/invoke interfaces), while passive buffers are read from and written
to directly by computational blocks and active buffers (using read/write
interfaces). Coordination functions are used to specify whether a given block is
executed in a passive or active fashion. Thus, coordination functions
specify how schedulers should manipulate the blocks when executing
the associated application graph.

Given a PAFG $F$, we represent the set of blocks (vertices) in $F$ by
$\mxblocks(F)$, and we define a {\em coordination function} of $F$ as one that
specifies for each $b \in \mxblocks(F)$ whether or not $b$ is to be executed in
an active or passive fashion.  More precisely, a coordination function is a
mapping $C: \mxblocks(F) \rightarrow \{\mxpassive, \mxactive\}$, where $C(b_c)
= \mxactive$ for every computational block $b_c \in \mxblocks(F)$, and $C(b_s)
= \mxpassive$ for every simple block $b_s \in \mxblocks(F)$.  We refer to
$C(b)$ as the {\em coordination type} of block $b$ with respect to $C$.
Computational blocks and simple blocks must be coordinated in an active
and passive fashion, respectively, and a coordination function just ``reminds
us'' of this.  On the other hand, a coordination function $C$ specifies for
each non-simple buffer block whether or not the block is to be executed in
a passive or active fashion (if we execute the PAFG based on $C$). 

A coordinated PAFG is an ordered pair
$Z = (F, C)$, where $F$ is a PAFG and $C$
is a coordination function for $F$.

\subsection{Alternating PAFGs}
\label{sec:alternating}

In this work, we are interested in a specific form of coordinated
PAFG,
which we refer to as an {\em alternating PAFG}. 
An alternating PAFG is defined to be 
a coordinated PAFG that is bipartite in terms of
the active blocks and passive blocks. More precisely,
an alternating PAFG $Z = (F, C)$ with $F = (V_f, E_f)$ is one that satisfies
$C(\mxsrc(e)) \neq C(\mxsnk(e)) \mathrm{for\ all\ } e \in E_{f}$.


A block in a PAFG is an {\em interface block} if it has no output edges or it
has no input edges.  The concept of coordinated PAFGs allows for the
possibility of interface blocks that are passive. However, we have not yet
experimented with the design of passive interface blocks.  Exploration into the
utility of passive interface blocks appears to be an interesting direction for
future work.

In our context, direct communication between pairs of
active blocks or pairs of passive blocks is ambiguous.
Intuitively, some form of buffer is needed 
to manage the flow of data between active blocks (just as
dataflow edges connect pairs of communicating actors
in dataflow graphs). Generalization of the developments
of this paper beyond alternating PAFGs is potentially another interesting
direction for future work.

\subsection{Direct PAFGs}
\label{sec:direct}

We propose a design methodology in which dataflow graphs
are converted into a kind of equivalent PAFG representation,
and then transformed so that some subset of
the active buffers is converted into passive coordination form.
In this section, we define the equivalent PAFG representation,
which we refer to as {\em direct PAFG} form, and
in Section~\ref{sec:passivization}, we define the process of transforming
active buffers into passive form.

Suppose that we are given a dataflow graph $G = (V, E)$.
For each edge $e \in E$, we define a corresponding passive buffer
$\rho(e)$. We denote the set of passive buffers defined
in this way as $P(G)$. Thus, $P(G) = \{\rho(e) \mid e \in E\}$.
Each $\rho(e) \in P(G)$ is a simple block (see Section~\ref{sec:components})
since it is defined in correspondence with a distinct 
dataflow graph edge $e$.

Similarly, for each $v \in V$, we define a corresponding block $\alpha(v)$.
Each $\alpha(v)$ is referred to as an {\em actor block} with corresponding
actor $v$. If $v$ is a computational actor, then $\alpha(v)$ is defined as a
computational block.  Otherwise, $\alpha(v)$ is defined as a non-simple 
buffer block.
For a given dataflow graph $G = (V, E)$, we define the set of all actor blocks
by $\mxallactors(G) = \{\alpha(v) \mid v \in V\}$ 

For each $z = \rho(e) \in P(G)$, we define the PAFG pairs $\mxinpair(z) =
(\alpha(\mxsrc(e)), z)$ and $\mxoutpair(z) = (z, \alpha(\mxsnk(e))$.  Recall
that PAFG pairs are ordered pairs of blocks, and actor blocks and passive
buffers both represent different types of blocks. Thus, $\mxinpair(z)$ and
$\mxoutpair(z)$ can correctly be referred to as PAFG pairs.  The sets of
all pairs defined in this way are represented by $\mxallin = \{\mxinpair(z)
\mid z \in P(G)\}$, and $\mxallout = \{\mxoutpair(z) \mid z \in P(G)\}$.

The {\em direct PAFG} representation of $G$ is 
a coordinated PAFG $Z_d = (F_d, C_d)$. The PAFG
$F_d = (V_d, E_d)$ is defined by
$V_d = (\mxallactors(G) \cup P(G))$, and $E_d = (\mxallin \cup \mxallout)$,
and the coordination function is defined by
$C_d(b) = \mxactive$ for every non-simple block $b$.

By construction, each edge in $Z_d$ connects a simple 
block to a computational block or an active buffer block.
Thus, a direct PAFG is always an alternating PAFG.

\begin{figure}[!t]
\centering
\includegraphics[width=2.0in]{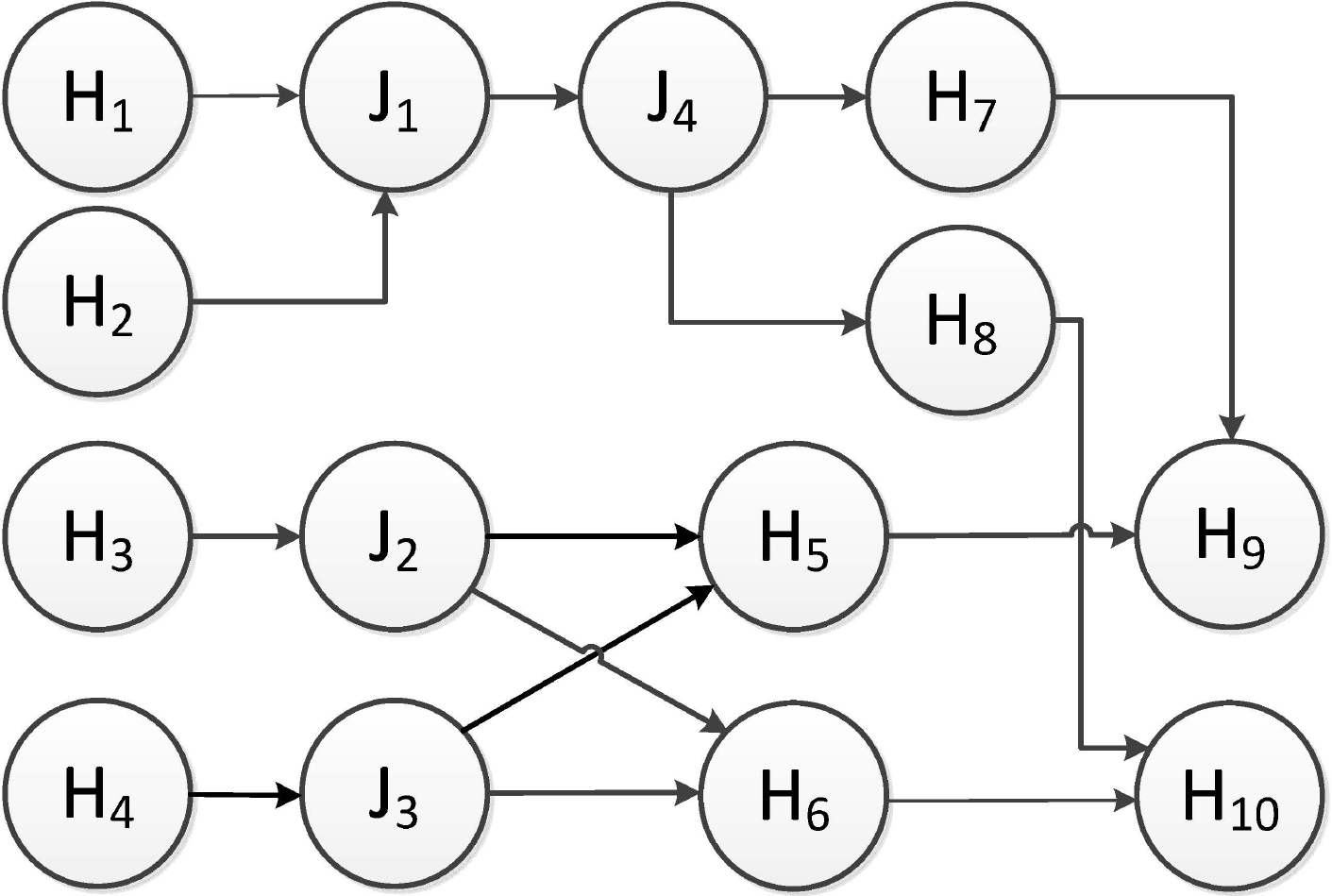}
\caption{A dataflow graph (application graph).}
\label{fig:pt-fg}
\end{figure}

\begin{figure}[!t]
\centering
\includegraphics[width=2.8in]{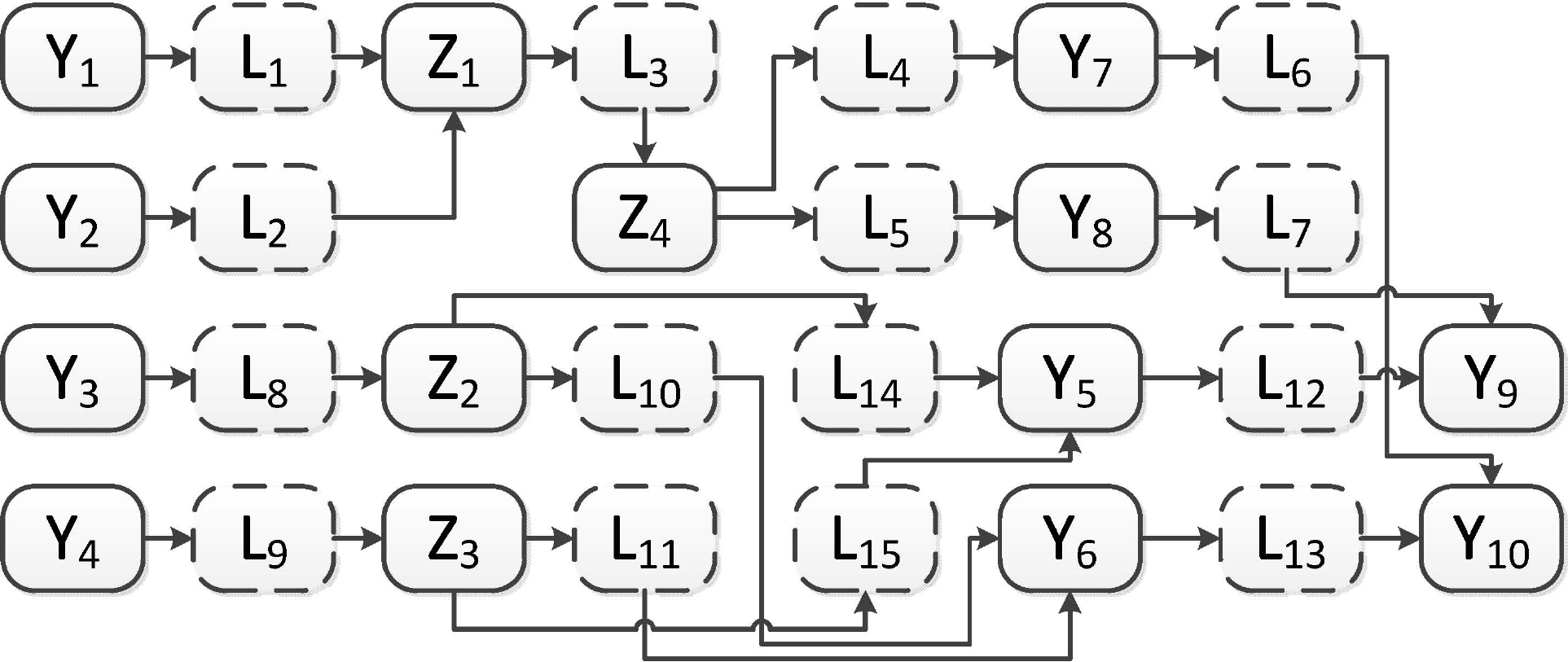}
\caption{The direct PAFG that is derived from the application graph
of Figure~\ref{fig:pt-fg}.}
\label{fig:pt-dfg}
\end{figure}

To illustrate key concepts introduced in this section,
Figure~\ref{fig:pt-fg} shows an example of a dataflow graph (application graph),
Figure~\ref{fig:pt-dfg} shows the direct PAFG
that results from this application graph,
and Table~\ref{tab:pt-cf-dfg} shows the coordination function
for the direct PAFG. In Figure~\ref{fig:pt-fg} and Figure~\ref{fig:pt-dfg}, each
$H_i$ is a computational actor,
each $J_i$ is a buffer actor,
each $Y_i$ corresponds to $H_i$,
each $Z_i$ corresponds to $J_i$,
and each $L_i$ is a simple passive buffer.

\begin{table}[!t]
\renewcommand{\arraystretch}{1}
\caption{Coordination function for the direct PAFG of Figure~\ref{fig:pt-dfg}.}
\label{tab:pt-cf-dfg}
\centering
\begin{tabular}{|c||c|}
\hline
Block ($B$) & Coordination type $C(B)$ \\
\hline
$Y_i, i = 1, 2, \ldots, 10$ & $\mxactive$ \\
\hline
$Z_j, j = 1, 2, \ldots, 4$ & $\mxactive$ \\
\hline
$L_k, k = 1, 2, \ldots, 15$ & $\mxpassive$ \\
\hline
\end{tabular}
\end{table}

As illustrated in Figure~\ref{fig:pt-fg} and Figure~\ref{fig:pt-dfg}, we use
the convention that dataflow graph actors are drawn with circles, PAFG blocks
are drawn with rectangles, and the borders of PAFG blocks are solid
or dashed based on whether the blocks are active or passive, respectively.

\subsection{Association between Dataflow Graphs and PAFGs}

Given a dataflow graph $G = (V, E)$ and
a PAFG $F = (V_f, E_f)$, we say that $G$ and $F$ are {\em associated} 
(each is associated with the other)
if each simple block $p$ in $F$ corresponds
to an edge $e$ in $G$ ($p = \rho(e)$),
and each non-simple block $q$ in $F$ corresponds
to an actor $a$ in $G$ ($q = \alpha(a)$).
By construction, the direct PAFG representation of 
a dataflow graph $G$ is always associated with $G$.

\section{Passivization Transformation}
\label{sec:passivization}

In the direct PAFG representation of a dataflow graph, all non-simple
buffer blocks are coordinated as active buffers. In this section, we define the
process of converting an active buffer to passive form.  This conversion
process is defined as a transformation process for alternating PAFGs ---
that is, a process that takes as input an alternating PAFG and produces as
output another alternating PAFG.

If $b$ and $c$ are adjacent blocks in a PAFG,
then we disallow coordination functions that
assign a passive form to both $b$ {\em and} $c$.
We refer to this as the {\em adjacent buffer coordination} 
({\em ABC}) restriction.
We impose the ABC restriction because we do not have any mechanism defined
for direct communication between two passive blocks.
Intuitively, communication between passive buffer blocks
``stalls'' because each is ``waiting'' for a read or write operation
to be initiated by the other.
It may be interesting as future work to investigate
communication mechanisms that allow
one to relax the ABC restriction.

Given an alternating PAFG $(F, C)$ and a block $b$ in $F$, we say that
$b$ is {\em simply surrounded} if all of its predecessors
and successors are simple passive buffers.
Formally, this means that $x$ is a simple passive buffer
for all $x \in (\mxpred(b) \cup \mxsucc(b))$.
For example, in Figure~\ref{fig:pt-dfg}, blocks $Z_1$ and  $Z_2$ are simply
surrounded, while blocks  $L_1$ and  $L_2$ are not.

Suppose that we have an alternating PAFG $Z_a = (F_a, C_a)$, where $F_a =
(V_a, E_a)$, and suppose we have an active buffer $\beta \in V_a$ that is
simply surrounded. Then we can perform the {\em passivization transformation}
of $Z_a$ with respect to $\beta$.  This transformation, which is the primary
contribution of this section, produces a new PAFG $Z_b = (F_b, C_b)$, $F_b
= (V_b, E_b)$.  The vertex set of $F_b$ is defined by 
the set difference $V_b = V_a - V_z$,
where $V_z = \mxpred(\beta) \cup \mxsucc(\beta)$.

To define the edge set $E_b$, we first define the sets $Y_p = \{y \in
\mxpred(x) \mid x \in \mxpred(\beta)\}$, and $Y_s = \{y \in
\mxsucc(x) \mid x \in \mxsucc(\beta)\}$.  Since $\beta$ is
simply surrounded, we have from the ABC restriction that all elements of
$Y_p$ and $Y_s$ are active blocks. Next, we construct the set $E_{\beta}$ of
PAFG pairs that are directed from members of $Y_p$ to $\beta$, or from $\beta$
to members of $Y_s$: $E_{\beta} = (\{(x,\beta) \mid x \in Y_p\} \cup \{(\beta,
y) \mid y \in Y_s\})$.  We also define the set of all input and output edges of
blocks that are adjacent to $\beta$: $E_r = \{e \in \mxout(x) \mid x \in V_z\}
\cup \{e \in \mxin(x) \mid x \in V_z\}$.  We can then define $E_b$ by $E_b =
((E_a - E_r) \cup E_{\beta})$.


The coordination function $C_b : V_b \rightarrow \{\mxpassive, \mxactive\}$ is
derived by changing the form of $\beta$, while ``copying'' the values from
$C_a$ for all other blocks in $V_b$: $C_b(\beta) = \mxpassive$, and $C_b(x) =
C_a(x)$ for all $x \in (V_b - \{\beta\})$.



To summarize, the passivization transformation with respect to a simply
surrounded active buffer $\beta$ involves the following steps: (1) changing the
form of $\beta$ from $\mxactive$ to $\mxpassive$; (2) removing all of the
predecessor and successor blocks of $\beta$ along with their input and output
edges; (3) adding edges that are directed to $\beta$ from each member of $Y_p$;
and (4) adding edges that are directed from $\beta$ to each member of $Y_s$.

The passivization transformation can be applied multiple
times, where in each application (transformation step)
after the first, the transformation is applied
on the graph that results from the previous step.

\begin{figure}[!t]
\centering
\includegraphics[width=2.3in]{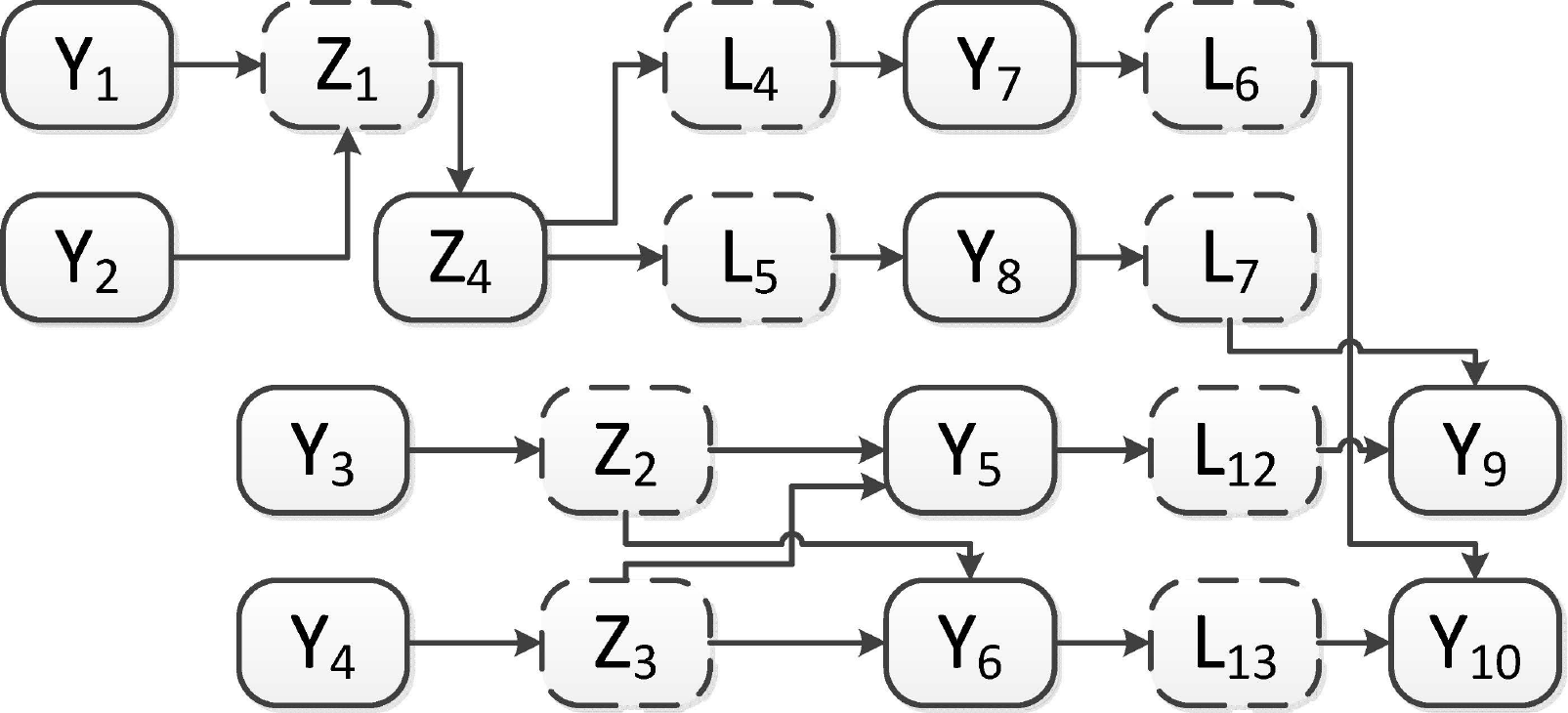}
\caption{Resulting PAFG after applying the passivization transformation.}
\label{fig:pt-fg-opt}
\end{figure}

For example, Figure~\ref{fig:pt-fg-opt} illustrates the PAFG that results after
applying the passivization transformation three times on the direct PAFG of
Figure~\ref{fig:pt-dfg}.  The transformation is applied with respect to $Z_1$,
$Z_2$, and then $Z_3$. 


\section{ Application Examples and Experiments}
\label{sec:experiments}

In this section, we present experiments on two relevant applications.  These
experiments demonstrate the utility of design optimization using PAFGs. In both
of these experiments, we carried out a sequence of passivization
transformations by hand, and implemented the original dataflow graph and the
optimized PAFG (derived through the transformations) using the lightweight
dataflow environment (LIDE)~\cite{lin2017x1}.  In this work, we have developed
extensions in LIDE to provide complete support for design and implementation
using PAFGs, including features that allow implementation and interfacing of
non-simple passive blocks.  The experiments for both applications are conducted
on an Intel Core i7-2600K Quad-core CPU running Ubuntu Linux 16.04 LTS, and
using GCC 5.4.0 for code compilation.

\subsection{Error Vector Magnitude Computation}

The error vector magnitude (EVM) is a figure of merit for signal quality in
communication systems.  EVM computation is an important application in
measurement and test equipment for communications.  For background on EVM
computation, we refer the reader to~\cite{schm2012x1}.

\begin{figure}[!t]
\centering
\includegraphics[width=2.0in]{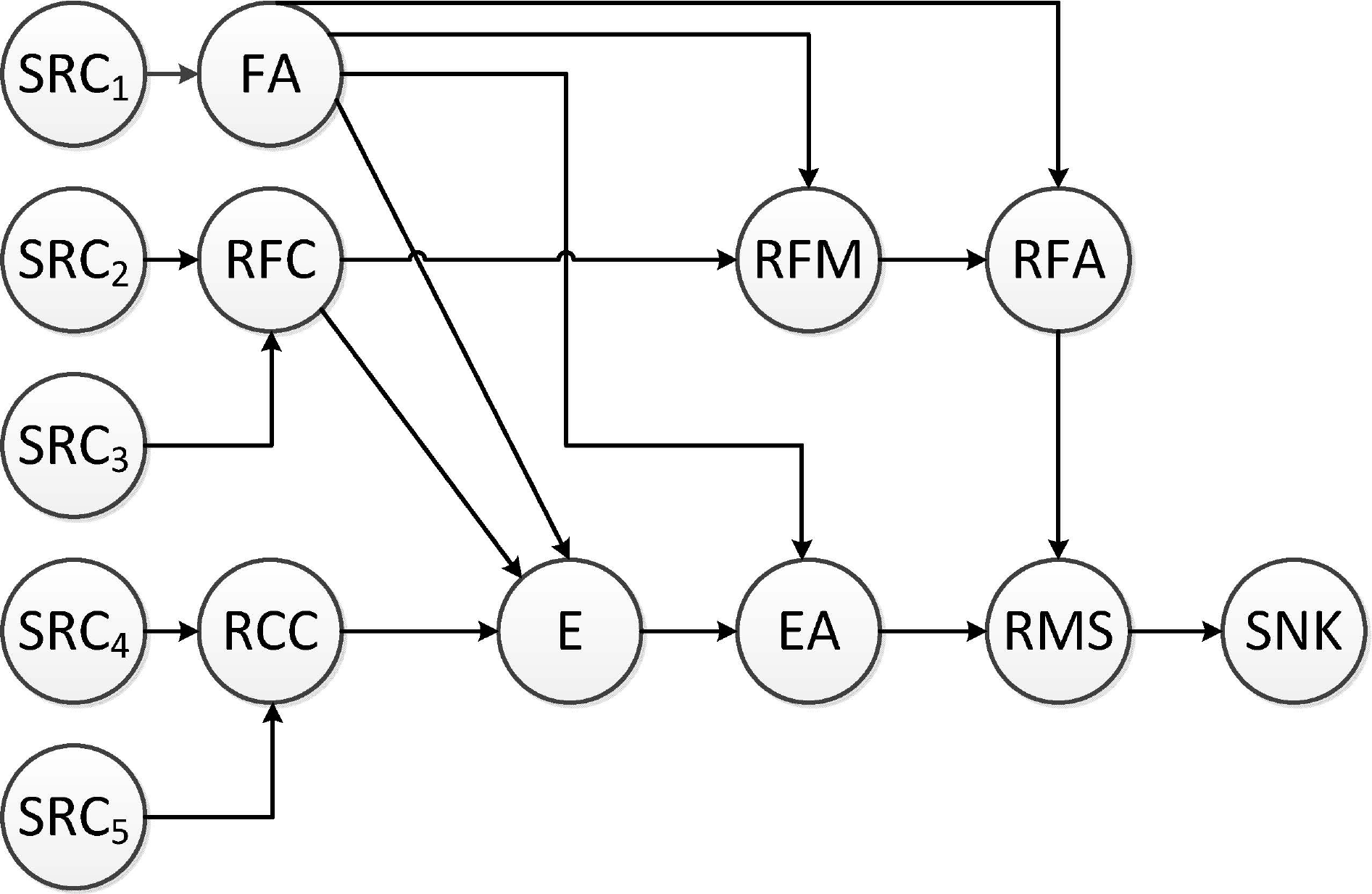}
\caption{Dataflow graph for EVM measurement.}
\label{fig:evm-df}
\end{figure}

A dataflow graph for measuring the EVM for a given reference signal and
received signal is shown in Figure~\ref{fig:evm-df}. This is a dynamic dataflow
graph modeled using CFDF semantics, as supported in LIDE. Here,
$\mathrm{SRC}_1$ provides on each $i$th firing the input data length for the
$i$th EVM computation. The actors $\mathrm{SRC}_2$ and $\mathrm{SRC}_3$ provide
the real and imaginary parts, respectively, of the reference signal; and
similarly, $\mathrm{SRC}_4$ and $\mathrm{SRC}_5$ provide the real and imaginary
parts of the received signal.  The actor $\mathrm{FA}$ is a fork actor (see
Section~\ref{sec:intro}), which broadcasts data to multiple output ports.  The
actors $\mathrm{RFC}$ and $\mathrm{RCC}$ are interleavers that interleave
corresponding pairs of input tokens so that the real and imaginary parts of
each signal sample are arranged in successive elements of the actors' output
streams.  The actors $E$ and $\mathrm{RFM}$ compute the error vector and
reference signal magnitude, respectively.  The actor $\mathrm{RMS}$ computes
the root mean square (RMS) $k_e$ of the error signal and the RMS $k_r$ of the
reference signal, and derives the EVM result as the ratio $k_e / k_r$.  The
actors $\mathrm{RFA}$ and $\mathrm{EA}$ compute the average magnitudes of the
reference and error signals, respectively.  The $\mathrm{SNK}$ actor represents
the output interface of the graph; in our experiments, we use a file writing
interface. 

We first derive a direct PAFG, which represents 
the implementation of the application graph (Figure~\ref{fig:evm-df})
using pure dataflow semantics. To the direct PAFG,
we apply the passivization transformation three times
with respect to the actors $\mathrm{FA}$, $\mathrm{RFC}$ and $\mathrm{RCC}$.
All three of these actors are simply-surrounded,
and can be implemented efficiently in passive form.

\begin{figure}[!t]
\centering
\includegraphics[width=2.6in]{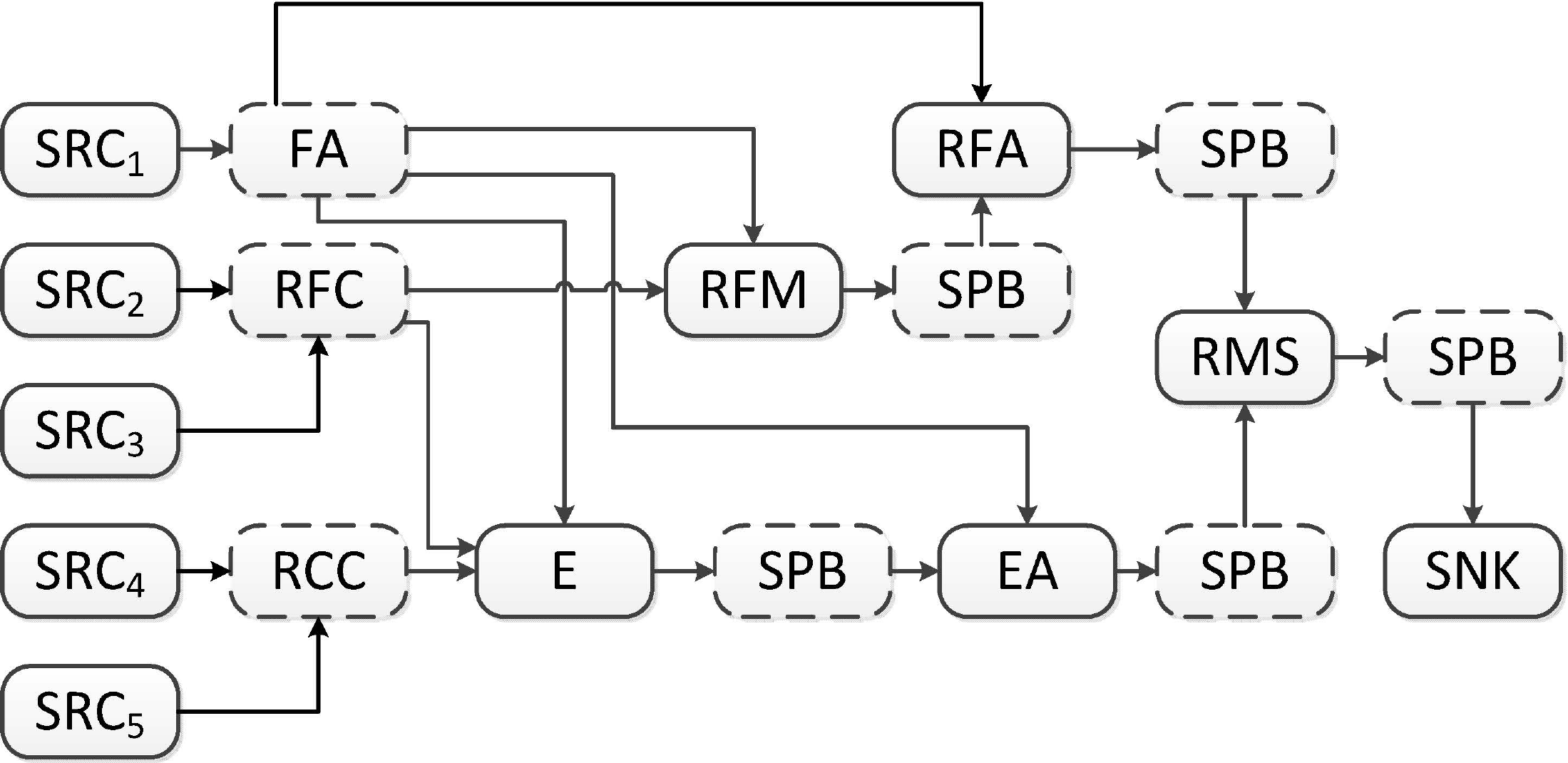}
\caption{Optimized PAFG for EVM measurement.}
\label{fig:evm-fg-b}
\end{figure}

The resulting optimized PAFG is illustrated in Figure~\ref{fig:evm-fg-b}.
We use a minor abuse of notation where non-simple blocks in the PAFG are
labeled with the same names as their corresponding actors in the
application graph. Blocks labeled as $\mathrm{SPB}$ represent
simple passive buffers.

Table~\ref{tab:evm-perf} compares the performance of the
direct and transformed PAFGs.
Through passivization, the throughput is 
improved by about 31.88\%, and the buffer memory requirement (BMR) is reduced
by about 25\%. We define the BMR of a PAFG $G$ as the 
total memory requirement for all passive blocks in $G$.



\begin{table}[h!t]
\caption{Results for the EVM application.}
\label{tab:evm-perf}
\centering
\begin{tabular}{|p{0.35\linewidth}||p{0.2\linewidth}|p{0.2\linewidth}|}
\hline
 & Throughput (samples/sec) & BMR (MB) \\

\hline
Direct PAFG & $7.93\times 10^{5}$ & 29.30\\
\hline
Optimized PAFG & $1.05\times 10^{6}$ & 21.97\\
\hline
\end{tabular}
\end{table}


\subsection{Jitter Measurement Application}

Jitter measurement is another important application for
real-time signal processing in communication systems.
In this section, we apply PAFG-based modeling
and optimization for a jitter measurement system design that
is presented in~\cite{liu2015x3}. For details on
this system design, including the dataflow model and the
constituent actors, we refer the reader to~\cite{liu2015x3}.

An important parameter in the jitter measurement system
is the {\em window size}, which determines the 
number of samples that are processed in a given dataflow
graph iteration. Larger window sizes in general improve the
throughput at the expense of a larger BMR~\cite{liu2015x3}.

\begin{figure}[!t]
\centering
\includegraphics[width=2.6in]{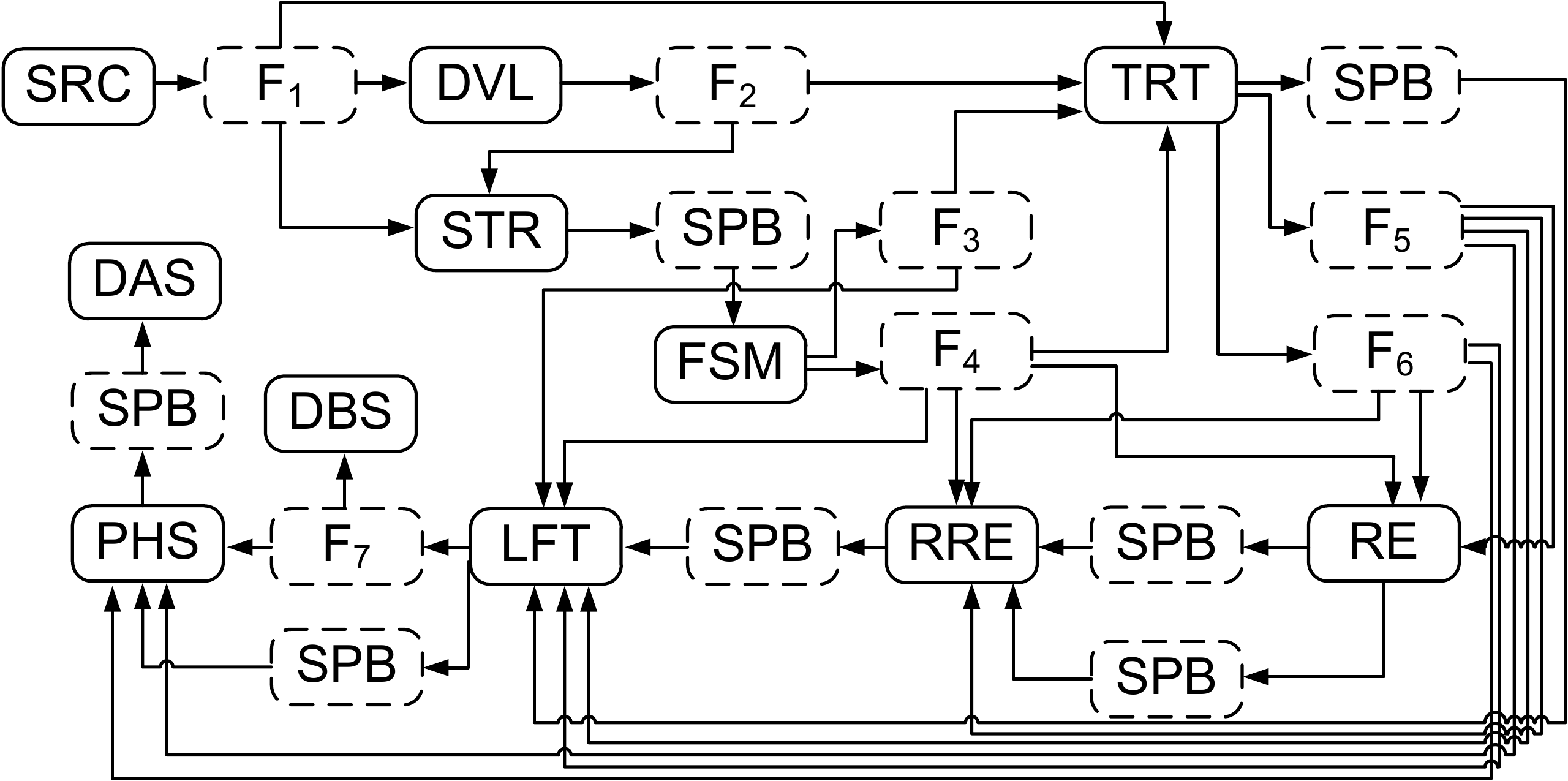}
\caption{Optimized PAFG for jitter measurement application.}
\label{fig:jitter-fg-b}
\end{figure}

Again, we first derive the direct PAFG and then
transform this into an optimized PAFG through
a sequence of passivization transformations. In this
transformation process, we convert the six fork actors in the design,
from active to passive buffer form.
The resulting optimized PAFG is illustrated in Figure~\ref{fig:jitter-fg-b}.
In this figure, the non-simple passive blocks corresponding
to the fork actors are denoted $F_1, F_2, \ldots, F_6$.

Table~\ref{tab:jitter-perf} shows the improvement measured from the optimized
PAFG compared to the direct PAFG for different window sizes.  From these
results, we see significant improvements delivered by the optimized PAFG in
terms of the trade-off between throughput and BMR.  For the optimized PAFG, the
BMR ranges from 0.38MB to 6.0MB for increasing window sizes, and the
throughput ranges from $1.9\times 10^{6}$ samples/sec to $3.1\times 10^{6}$
samples/sec.

\begin{table}[h!t]
\caption{Results for the jitter measurement application.}
\label{tab:jitter-perf}
\centering
\begin{tabular}{|p{0.18\linewidth}||p{0.09\linewidth}|p{0.09\linewidth}|p{0.09\linewidth}|p{0.09\linewidth}|p{0.09\linewidth}|}
\hline
Window size & 16,384 & 32,768 & 65,536 & 131,072 & 262,144 \\
\hline
Throughput & $11\%$ & $7.0\%$ & $7.7\%$ & $6.5\%$ & $7.0\%$\\
\hline
BMR & $60\%$ & $60\%$ &$60\%$ &$60\%$ &$60\%$ \\
\hline
\end{tabular}
\end{table}

\section{Conclusion and Future Work}
\label{sec:conclusion}

In this paper, we have introduced passive-active flowgraphs (PAFGs) as a model
of computation that complements dataflow models for design and implementation
of signal processing systems. PAFGs generalize the concept of dataflow edges
into multi-input, multi-output components that are called ``passive blocks''.
PAFGs provide a new approach to integrating designer-specified memory
management optimization systematically into the framework of dataflow-based
design and implementation.  In addition to presenting details of the PAFG model
of computation, we have introduced the passivization transformation, which can
be used iteratively to derive progressively more efficient PAFGs.  We have
demonstrated the utility of PAFGs and the passivization transformation on two
important signal processing applications. Useful directions for future work
include automating the equivalence checking between active and passive versions
of a given actor, and the generalization of relevant methods in this paper
beyond alternating PAFGs.

\section{Acknowledgments}
\label{sec:ack}

This research was supported in part by 
the U.S.~National Science Foundation.

\bibliographystyle{IEEEtran}
\bibliography{IEEEabrv,refs}
%



\end{document}